\documentclass[a4paper]{jpconf}
\usepackage{graphicx}
\usepackage{lineno}
\usepackage{hyperref}
\usepackage{color}
\usepackage{url}

\begin{document}

\title{Future Computing Platforms for Science in a Power Constrained Era}

\author{David Abdurachmanov$^1$, Peter Elmer$^2$, Giulio
Eulisse$^1$, Robert Knight$^3$}

\address{$^1$ Fermilab, Batavia, IL 60510, USA}
\address{$^2$ Department of Physics, Princeton University, Princeton, NJ 08540,
USA}
\address{$^3$ Research Computing, Office of Information Technology, Princeton
University, Princeton, NJ, 08540, USA}

\ead{Giulio.Eulisse@cern.ch}

\begin{abstract}
Power consumption will be a key constraint on the future growth of Distributed
High Throughput Computing (DHTC) as used by High Energy Physics (HEP). This
makes performance-per-watt a crucial metric for selecting cost-efficient
computing solutions. For this paper, we have done a wide survey of current and
emerging architectures becoming available on the market including x86-64
variants, ARMv7 32-bit, ARMv8 64-bit, Many-Core and GPU solutions, as well as
newer System-on-Chip (SoC) solutions. We compare performance and energy
efficiency using an evolving set of standardized HEP-related benchmarks and
power measurement techniques we have been developing. We evaluate the potential
for use of such computing solutions in the context of DHTC systems, such as the
Worldwide LHC Computing Grid (WLCG).
\end{abstract}

\section{Introduction and Motivation} The data produced by the four experiments
at the Large Hadron Collider (LHC)~\cite{LHCPAPER} or similar High Energy
Physics (HEP)  experiments requires a significant amount of human and computing
resources which  cannot be provided by research institute or even country. For
this reasons the  various parties involved created the Worldwide LHC Computing
Grid (WLCG) in  order to tackle the data processing challenges posed by such a
large amount of data. The WLGC consists of a highly federated union of computing
centers sparse in 40 countries and it represents an admirable example of
international organization. The Compact Muon Solenoid (CMS) 
experiment~\cite{CMSDET} at the LHC  alone uses the order of 100,000 x86\_64 cores for its
data analysis and similar  happens for the other general purpose LHC experiment,
ATLAS.  Moreover, as the LHC and the experiments will undergo planned luminosity
upgrades over the next 15 years~\cite{HLLHC}, its dataset size will increase of
2-3 order of magnitude, which will require additional efforts to increase its
processing capacity.

In this paper we continue our ongoing effort to explore the performance and
viability of various computing platforms which we consider some of the most
likely components of tomorrow data centers. We asses their performance based on
both synthetic benchmarks and realistic workflows used in production at CMS.

\section{Considered platforms}
For this study we have selected five major product lines:

\begin{itemize}

\item \textbf{Intel Xeon}: Intel is the leading CPU manufacturer in the world.
Its Xeon product line is its flagship brand and provides solutions for all
ranges of computing centers. While the Xeon brand is usually associated to
highly performant, architecturally advanced CPUs, over the years Intel made sure
to take into account power efficiency needs. This is reflected by the inclusion
of features, like SpeedStep and TurboBoost, specifically aimed at a more power
efficient usage or RAPL which is used to monitor CPU power usage itself.

\item \textbf{APM XGene}: ARM architecture based products are, when considered
as a single entity, market leader for low power CPUs, in particular due to their
wide spread adoption in a wide range of consumer electronics product like
cellular phones and tablets. Thanks to the economy of scale of cell phones aims
to become a serious player in the server market, in particular Applied Micro
(APM) X-Gene product line is one of the first attempts at providing a 64-bit
ARMv8 chip which is suitable for the low power, high density server market. We
have already detailed the main difficulties sustained to port to such an
architecture in a preceding work~\cite{ACAT2014ARCH}.

\item \textbf{Intel Atom}: Intel Atom architecture is Intel solution for the
power efficient market. It consists of a standard x86\_64 core, where particular
trade-offs have been made to reduce complexity, sacrificing performance for
power efficient. E.g. Atom processors has a simpler, in order, architecture
without HyperThreading, limited vector units and cache subsystem when compared
to a standard Xeon. Atom has however the advantage that code compiled with
standard optimizations (i.e. \textit{-O2}) runs unmodified on it.

\item \textbf{IBM POWER8}: The POWER8 is the latest incarnation of the POWER
product line which is the evolution of the old PowerPC. While the latter was the
result of a strategic alliance between Apple, Motorola and IBM, the former,
initially an IBM only brand, is now managed by an industry consortium, Open
POWER Foundation~\cite{OPENPOWERFOUNDATION}, which includes big players like
Google, NVidia, Tyan. Compared to the past, the POWER8 includes efforts to
simplify platform ports from x86\_64 and has the usual focus on highly threaded
workloads, by providing an high number of cores.

\item \textbf{Intel Xeon Phi}: the Xeon Phi is Intel answer to the GPGPU market,
which in recent years has dominated the scene of High Performance Computing when
the problem to solve maps well on a many-core architecture like the one of GPUs.
The Phi consist of a very high number of simplified x86\_64 cores, running at a
relatively low frequency, which however have a large vector unit. The advantage
touted by Intel for such an architecture is that being real x86\_64 cores, the
porting effort is lower when compared to writing software for a GPU.

\end{itemize}

\section{Test Environments for Power and Performance Measurements}

We now describe the test environments we have used to do power and performance
measurements for two Intel Xeon processors, belonging respectively to the Sandy
Bridge generation  and to the Haswell one, an APM ARMv8 64-bit X-Gene1
Server-on-Chip, an Intel Xeon Phi coprocessor, an Intel Atom processor of
Avoton class, an IBM POWER8 processor.

\subsection{Hardware setup}
Table \ref{models} details some of the general details of those processors, in
which one can already spot the almost two years advantage which Intel has in
terms of fabrication process, when compared to ARM based solutions.

\begin{table}[ht]
\caption{\label{models}CPU models used}
\begin{center}
    \begin{tabular}{llllll}
        \br
            Name             & Vendor & Model       & Year    & Fab   & Process  \\
        \mr
            Xeon SandyBridge & Intel  & E5-2650     & Q1/12   & Intel & 32nm \\
            Xeon Haswell     & Intel  & E5-2699     & Q3/14   & Intel & 22nm \\
            X-Gene1          & APM    & APM883408   & Q3/13   & TMSC  & 40nm \\
            Atom             & Intel  & C2750       & Q3/13   & Intel & 22nm \\
            POWER8           & IBM    & IBM8247-22L & Late 13 & IBM   & 22nm \\
            Xeon Phi         & Intel  & KNC7100     & Q2/14   & Intel & 22nm \\
        \br
    \end{tabular}
\end{center}
\end{table}

In table \ref{specs} actual specifications of the benchmarked models are shown.
As one can see from such table there are two obvious tradeoffs which have been
made on low power chips (i.e. X-Gene1 and Atom), removing the
HyperThreading-like subsystem and keeping the number of cores down. On the other
hand the POWER8 went for a much higher number of threads, which is explained by
the fact it's marketed at very high end servers, with highly parallel workloads,
e.g. web servers. As already mentioned the Xeon have the ability to scale up
single core performance via the so called TurboBoost feature that allows
increasing the core frequency when only a few of the cores are being utilized
(reported in parenthesis).

\begin{table}[h]
\caption{\label{specs}Silicon chips specifications}
\begin{center}
    \begin{tabular}{lrrl}
        \br
                             & \# Cores & \# Threads & Frequency (GHz) \\
        \mr
            Xeon SandyBridge & 8       & 16        & 2.0 (2.8)       \\
            Xeon Haswell     & 18      & 36        & 2.3 (3.6)       \\
            X-Gene1          & 8       & 8         & 2.4             \\
            Atom             & 8       & 8         & 2.4             \\
            POWER8           & 10      & 80        & 3.4             \\
            Xeon Phi         & 61      & 252       & 1.21            \\
        \br
    \end{tabular}
\end{center}
\end{table}

\subsection{Benchmarks setup}

We selected three benchmarks for our study, which are commonly used in HEP and
in particular in CMS to determine the performance of a machine.

\begin{itemize}

\item \textbf{ParFullCMS}: a standalone Geant4~\cite{GEANT4} simulation using a
geometry similar to the one used in production by CMS, with simplified physics.
This benchmark has been recently adopted across LHC experiments for
multithreading studies, since on traditional architectures it already scales to a
large number of threads. For our benchmark we used ParFullCMS which
ships with Geant4.10.3.

\item \textbf{CMSSW RECO}: the reconstruction of 100 $t\overline{t}$ events with
pileup 35 at 25ns, using CMS Offline Software (CMSSW). This is considered a
standard candle to measure the performance of CMS reconstruction as it
guarantees good coverage of the code associated to all of reconstructed objects.
For our benchmark we used the development branch of CMSSW, 7.5.x, as of April
1st 2015.

\item \textbf{HEPSPEC06}: a HEP driven benchmark suite, widely used as it should
scale as HEP specific workloads~\cite{Michelotto2010}. In particular it is used
for LCG site pledges and therefore purchases.

\end{itemize}

All the benchmarks were compiled using the latest version of GCC 4.9.x available
on a given platform. The only notable exception to this was the Phi benchmarks,
which used the Intel Compiler (ICC 15.0.2) given the lack of support for vector
units of GCC. Since we wanted to be as close as possible to the production setup
used by CMS when running on the grid, we switched on conservative optimization
flags (i.e. \textit{-O2}) and we compiled code targeting vector units, we did
not use any platform specific optimization or aggressive optimization option
(e.g. \textit{-ffast-math}). In particular we used the same exact binaries for
Xeon and Atom as platform compatibility is considered by Intel one of Atom
advantages over other low power architectures in an heterogeneous environment,
like the Grid. Of course, this limits the capabilities of the Xeon, which for
example supports much advanced vector instructions (AVX and AVX2) compared to
Atom (which only supports SSE4) or the other processors, however we deemed that
this setup is closer to what is actually done today in the Grid where usually
there is no selection of optimized binaries for more recent CPU architecture.
Enabling this and benchmarking results is of course interesting, but outside the
scope of this paper.

Given Xeon SandyBridge is right now one of the most popular CPUs on the Grid,
and in order to simplify comparisons between different benchmarks, we normalized
all the results to those of the same benchmark running on in single core mode on
our SandyBridge test system.

\section{Methodology}

In the ParFullCMS case we ran a total of 1220 events, subdividing the workload
between an increasing number of threads, up to the number of hardware threads
the CPU had (in case HyperThreading like mechanism was available) or double the
number of hardware cores (in case the CPU did not support hardware). The
throughput of the CPU was measured by dividing the number of events by the
number of seconds spent in the event processing loop, as reported by ParFullCMS.
This number is an underestimation of the performance of a CPU, but it's deemed
to be a good enough approximation, especially for a low number of threads.

In the CMSSW case we started an increasing number of processes, using the same
logic as before to determine the maximum number of processes. Each process is
running the same amount of events and in the end we calculate the total throughput
by summing the throughput of each job. This ends up overestimating the actual
throughput of short running jobs, but it's a closer match to the actual
production case.

For the HEPSPEC06 case we started the benchmark specifying the relevant option
to run it multithreaded and used the number reported at the end as indication of
the performance of the system.

We have previously detailed our methodology on how to measure CPU power 
utilization~\cite{ACAT2014IGPROF,ACAT2014ARCH}. On-chip sensors (Running
Average Power Limit -- RAPL) were used to measure silicon level power
consumption for Xeon CPUs. Similar on-board sensors were available for X-Gene1
and Xeon Phi products. HP Moonshot chassis was equipped with power consumption
reporting, but external power distribution unit (PDU) was used instead as it
provided data logging capability. Only a single power supply unit (PSU) was used
in HP Moonshot systems while conducting experiments. Unfortunately we were not
able to get power consumption measurements for the POWER8 CPU or the box. Atom
systems did not provide RAPL reporting thus we used a full node power
consumption measurements.

\section{Results}

\subsection{Raw performance}
\begin{figure}[ht]
\centering
\includegraphics[width=12cm]{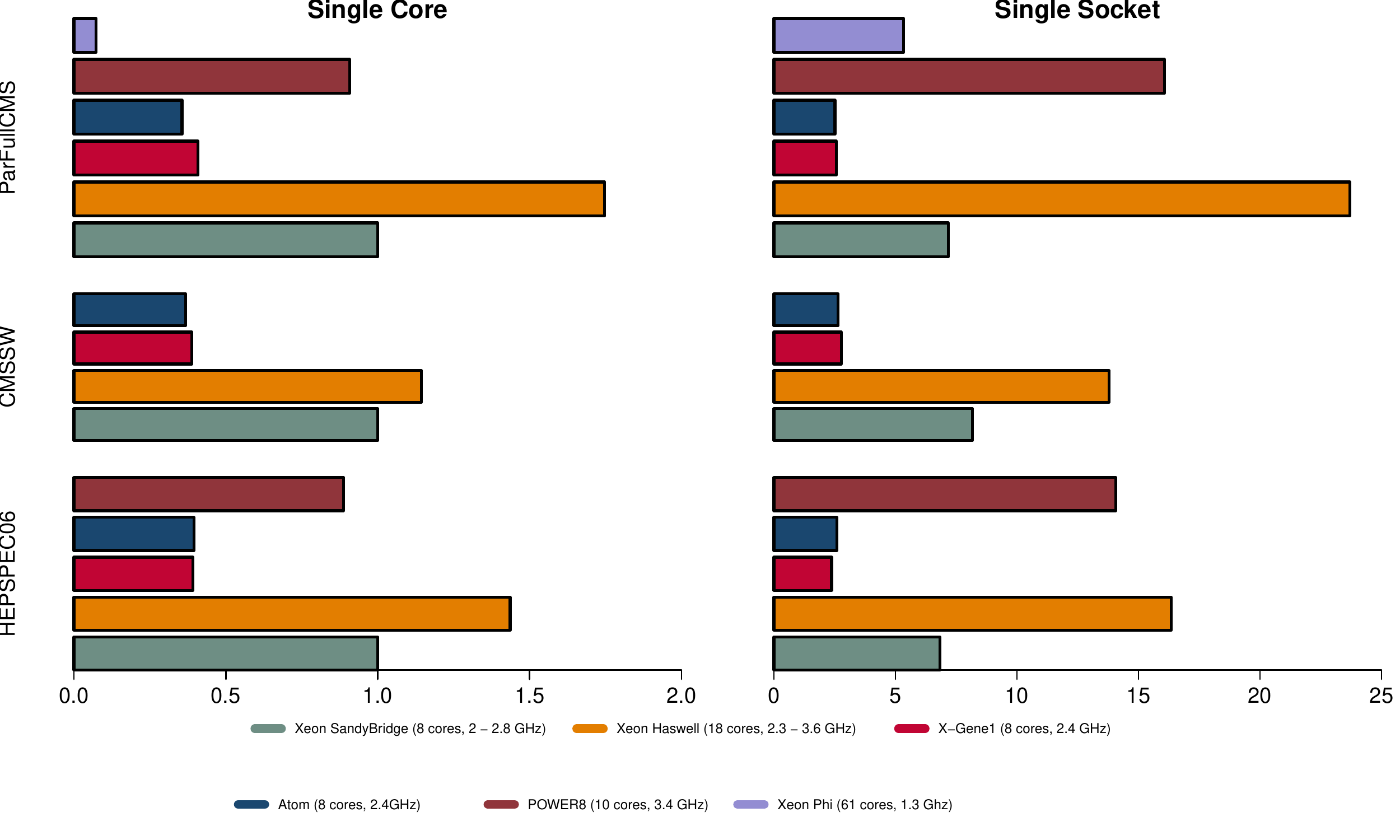}
\caption{\label{fig:rawPerformance}Raw performance results}
\end{figure}

In figure ~\ref{fig:rawPerformance} we show the results of all our benchmarks.
As anticipated, we can clearly identify two class of performance, Xeons and
POWER8 on one side and X-Gene and Atom on the other side.

\begin{figure}[ht]
  \centering
  \begin{minipage}{7.0cm}
    \includegraphics[width=7.0cm]{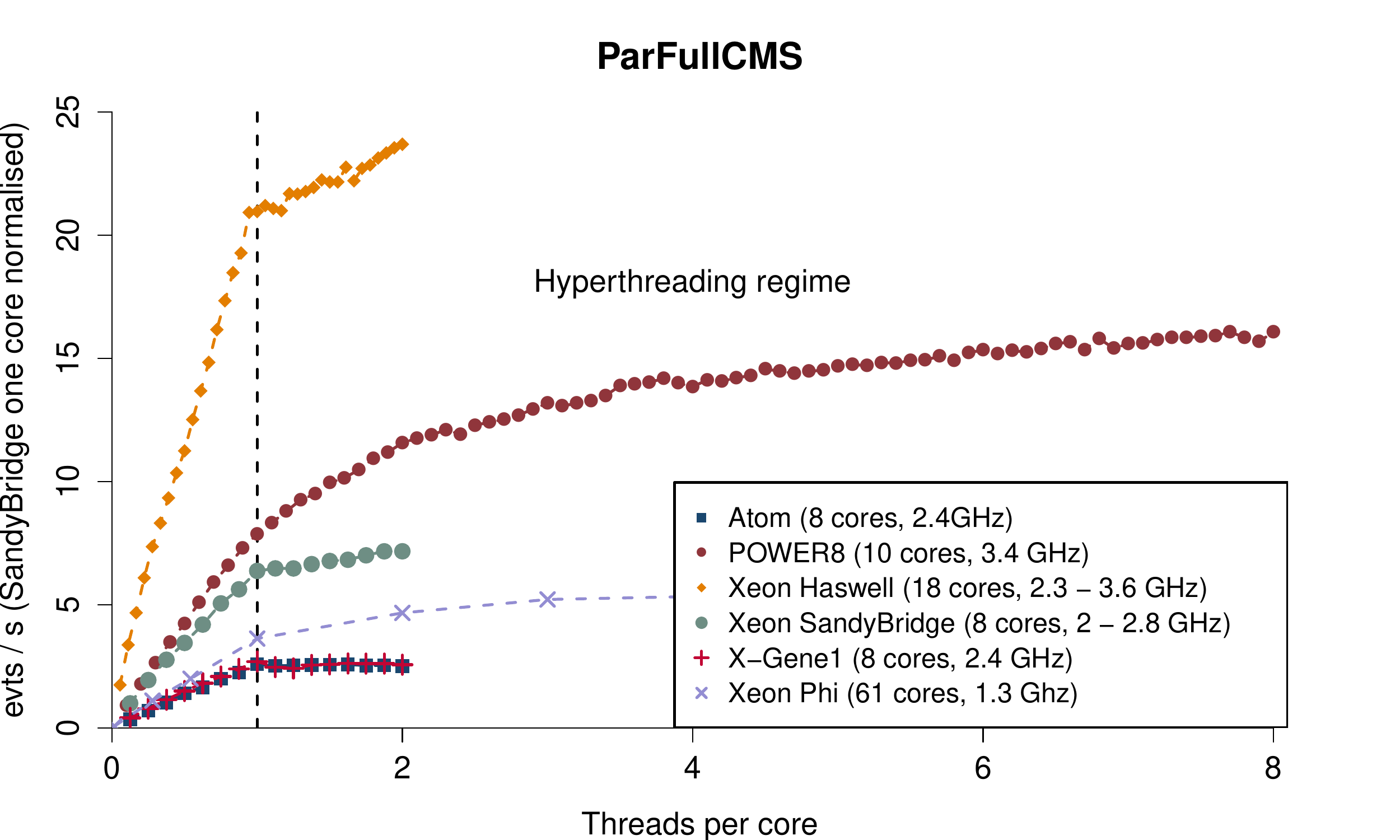}
    \caption{\label{fig:pfcScalability}Performance scalability}
  \end{minipage}
  \hspace{0.5cm}
  \begin{minipage}{7.0cm}
    \includegraphics[width=7.0cm]{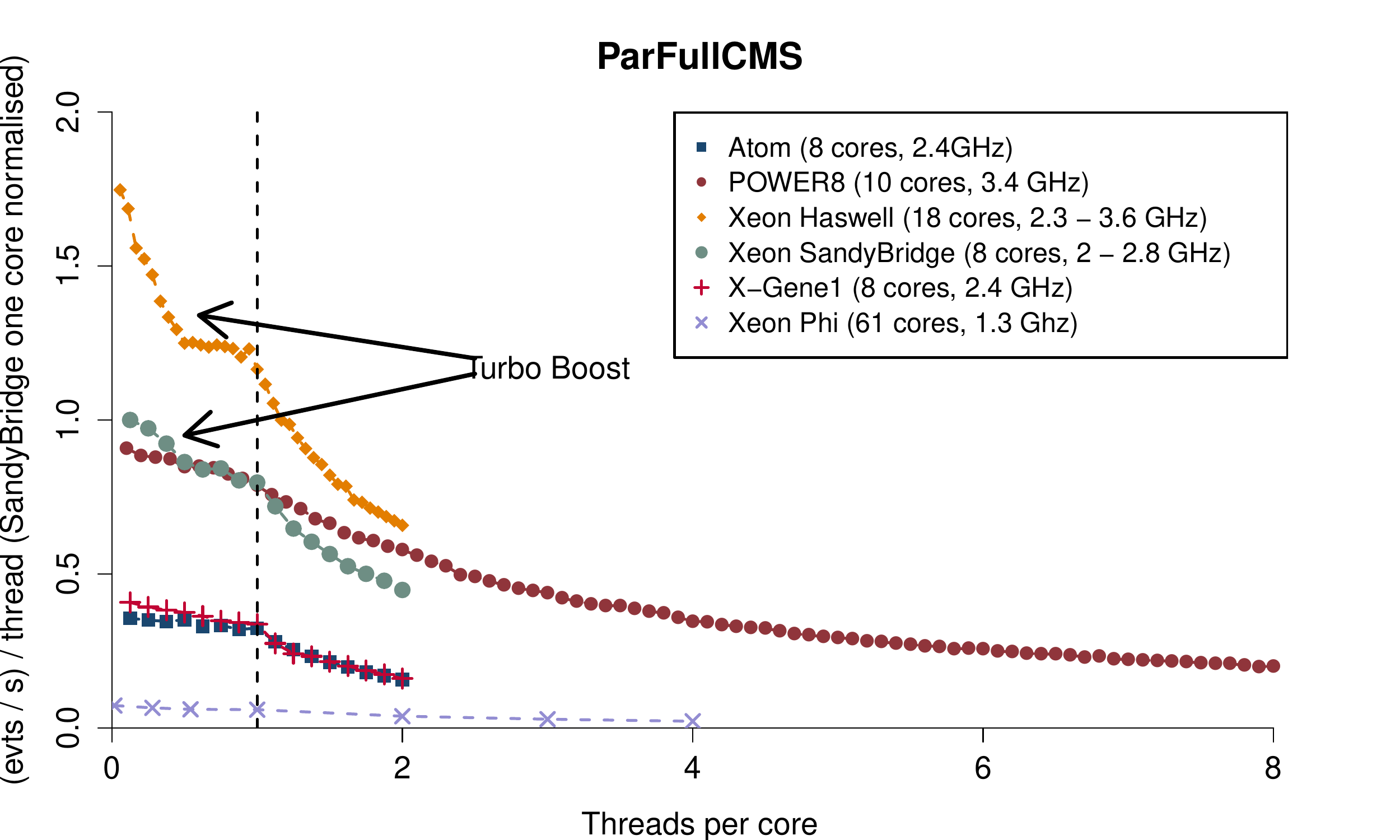}
    \caption{\label{fig:pfcTurboost}Per core performance}
  \end{minipage}
\end{figure}

By looking at figure ~\ref{fig:pfcScalability}, one can immediately see which
CPU has HyperThreading (Xeon, POWER8) and which does not (Atom, X-Gene1). In
particular we see how POWER8 hardware threading scales better than the others,
thanks to the eight hardware threads per core, but it's far from perfect scaling
when in HyperThreading regime. We attribute the disappointing performance of
Xeon Phi to the fact that the benchmark is a direct port of the multithreaded
application without any specific Phi improvement and optimization. We
nevertheless decided to include the results in this comparison to point out how
Xeon Phi (and to a lower degree POWER8) do need a non negligible code
optimization effort in order to perform to their maximum.

Similarly in figure ~\ref{fig:pfcTurboost} we have plotted the per core
performance, which immediately highlights how TurboBoost provides Xeon
additional performance with a lower CPU usage. This highlights the importance
of benchmarking a modern CPU with a load close to the average production one,
since single process benchmarks will always overestimate performance.

In all our benchmark the new Intel Xeon is shown to be the best overall
performer, both for single thread and fully loaded socket tests. In the fully
loaded case, the only contender seems to be the POWER8, with very close HEPSPEC6
absolute results, if one does not consider the different operating frequencies.

\subsection{Power efficiency}

\begin{figure}[ht]
\centering
\begin{minipage}{7.5cm}
\includegraphics[width=7.5cm]{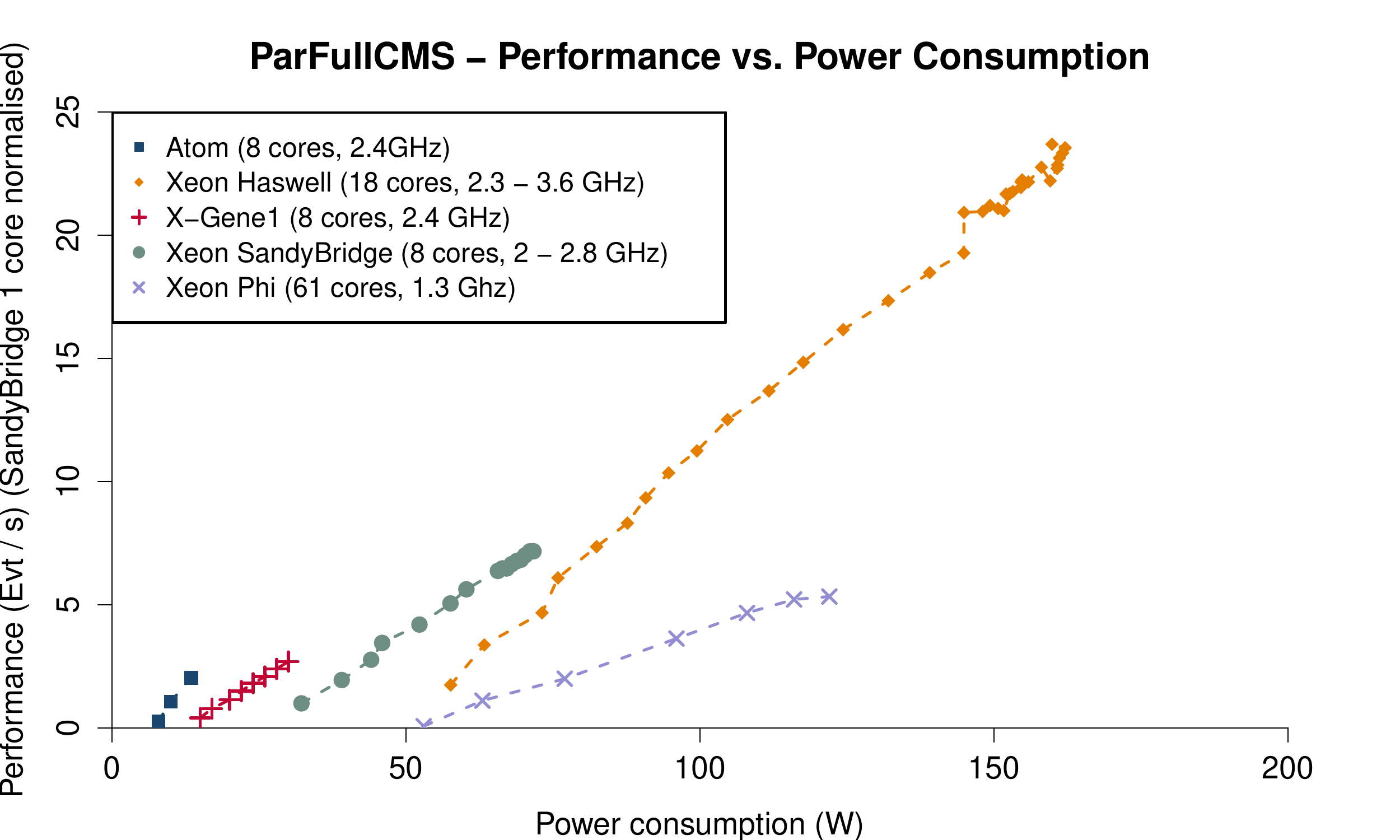}
\caption{\label{fig:performanceVsPower}Performance per power consumed}
\end{minipage}
\hspace{0.3cm}
\begin{minipage}{7.5cm}
\includegraphics[width=7.5cm]{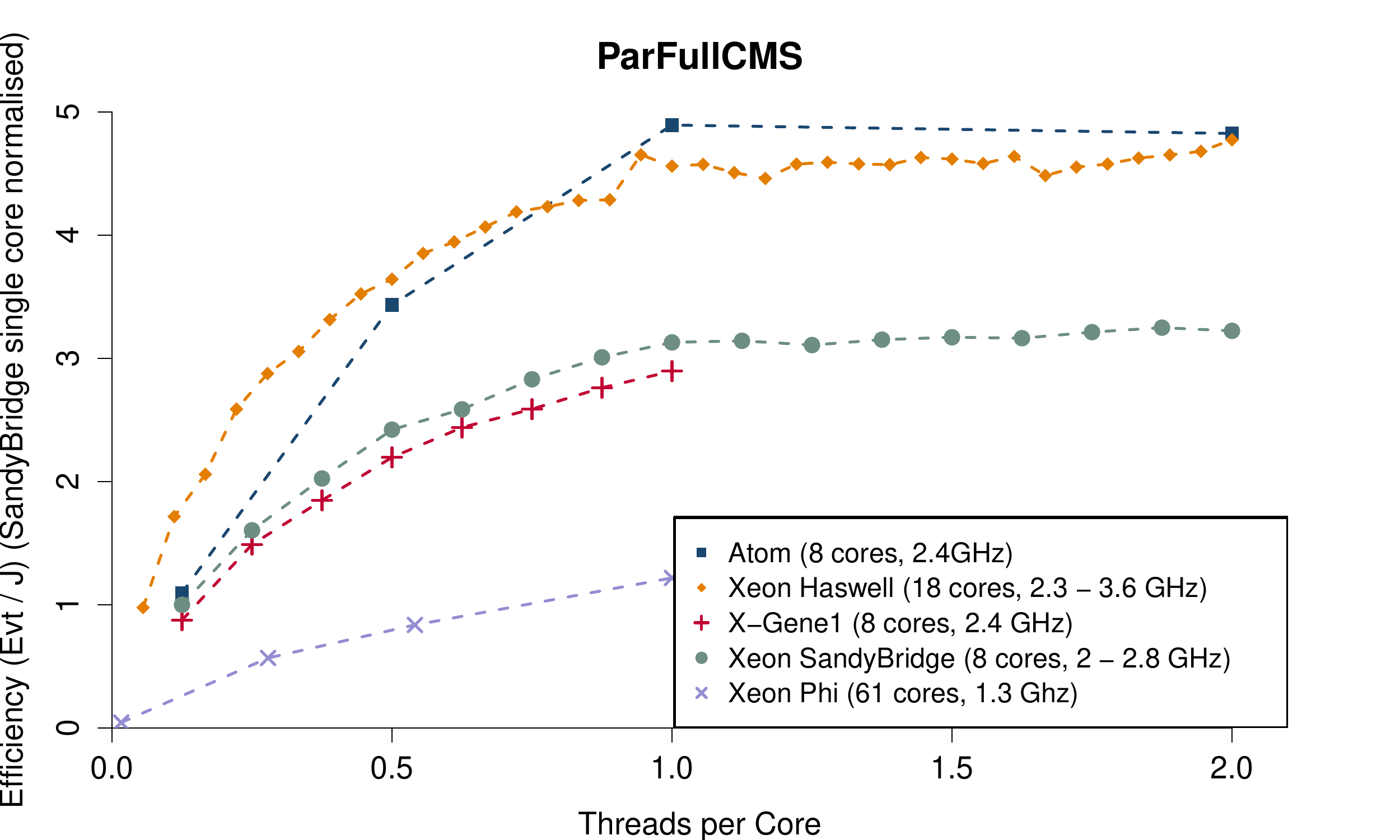}
\caption{\label{fig:powerEfficiency}Energy efficiency scalability}
\end{minipage}
\end{figure}

\begin{figure}[ht]
\centering
\includegraphics[width=12cm]{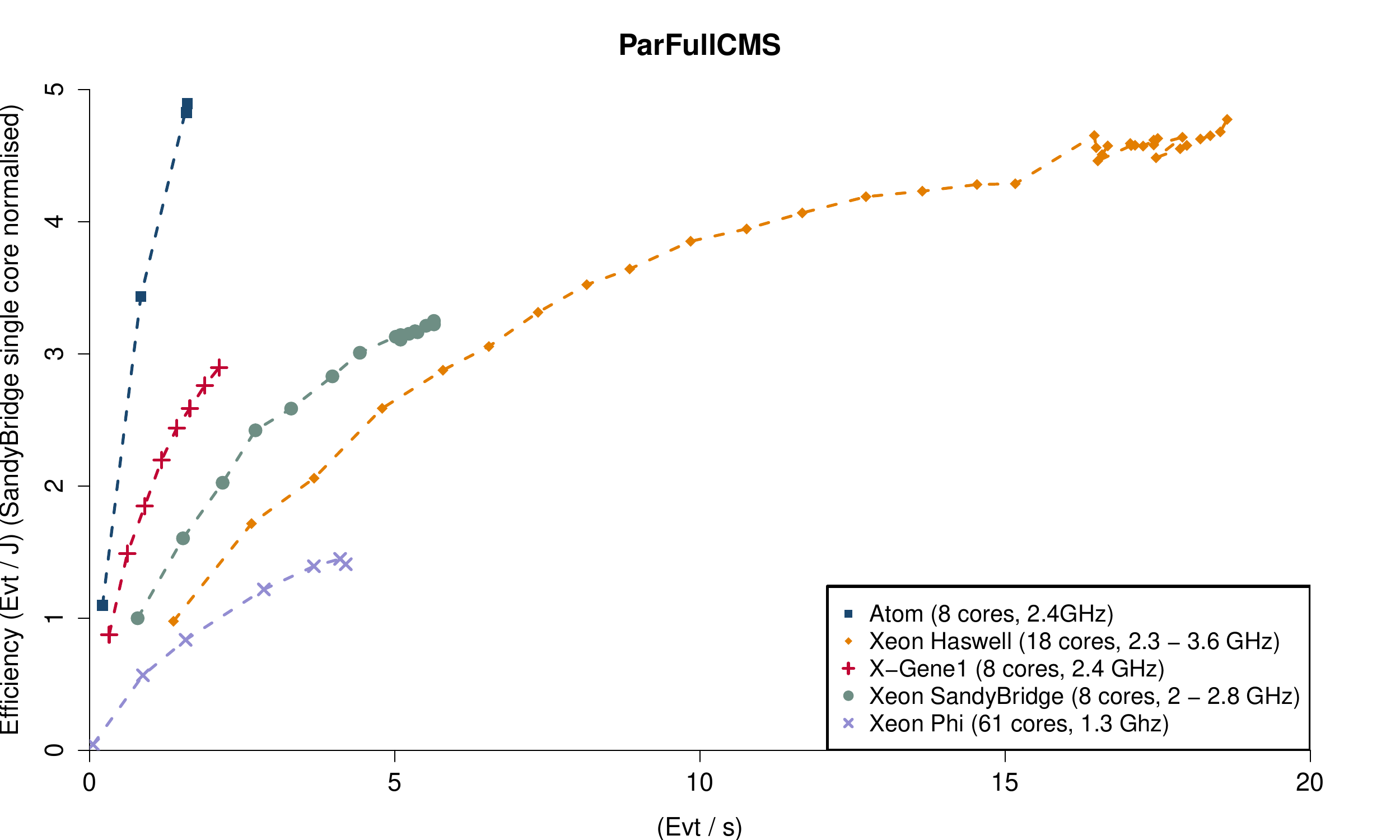}
\caption{\label{fig:efficiencyVsPerformance}Power efficiency at given performance}
\end{figure}

As we said, raw performance comes with a price tag in terms of power consumption
as it can be seen in figure~\ref{fig:performanceVsPower}. While Haswell performs
extremely well, it's also the one that requires most of the power to run. The
actual efficiency of each system can be better evaluated looking at figure
~\ref{fig:powerEfficiency} which clearly shows that Haswell and Atom are
in the same league. While in the past we reported X-Gene1 as a possible
contender, it's also clear that Intel is not sitting idle and without continuous
effort to steadily  follow an improvement roadmap, newer generation of Intel
chips quickly advance not only in terms of raw performance but power efficiency
in general. One of the possible metrics to decide how to select between
similarly efficient CPUs can be fount in figure
~\ref{fig:efficiencyVsPerformance} which provides power efficiency as a function
of performance. Fixing the wanted / required performance level immediately gives
the platform which performs better in terms of power efficiency.

\subsection{Box to box comparison}

\begin{figure}[ht]
\centering
\includegraphics[width=12cm]{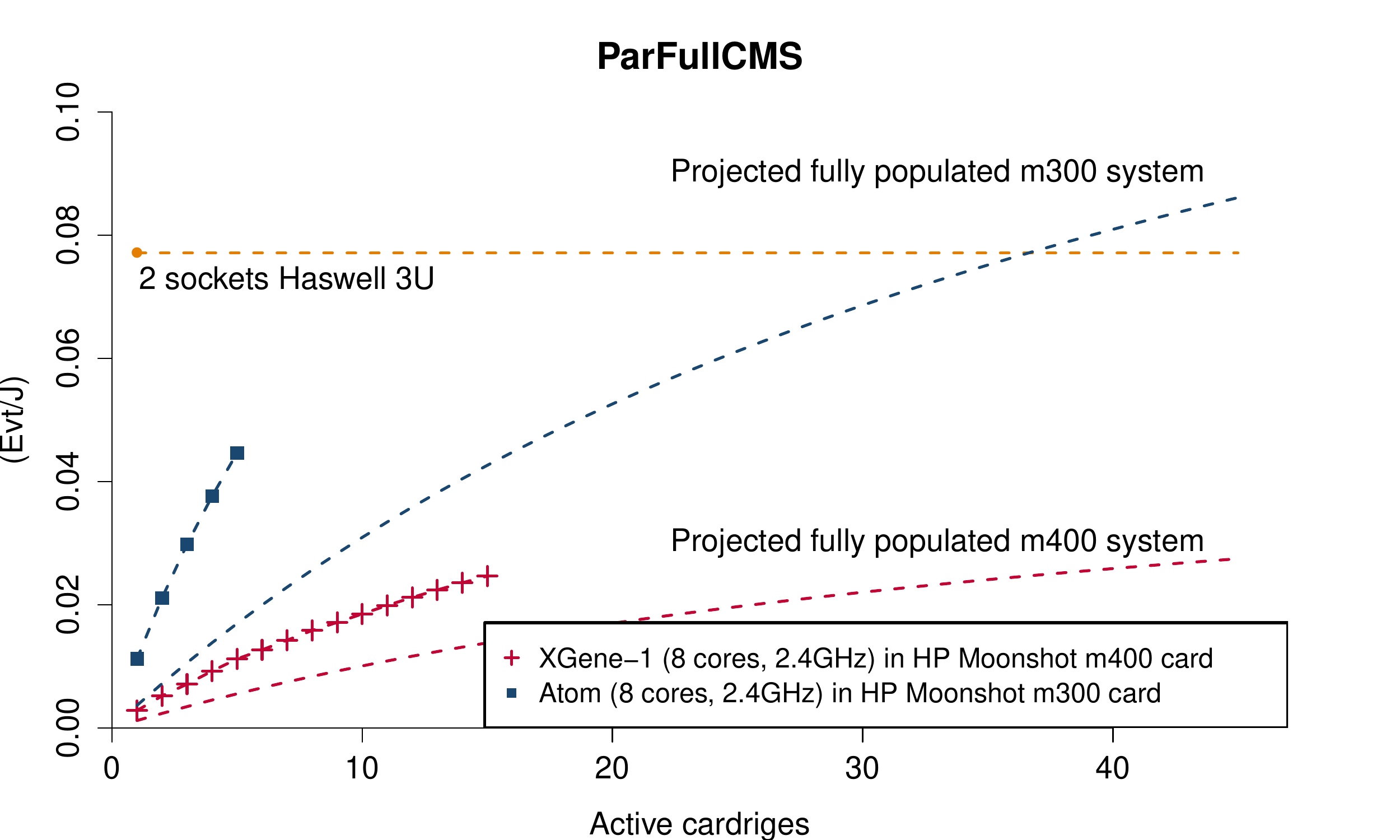}
\caption{\label{fig:box2box}Box to box comparison for HP Moonshot system}
\end{figure}

For our tests we had the privilege to try out an HP Moonshot chassis, which has
the peculiarity of supporting cartridges with Atom (the m300 model) and X-Gene1
(the m400 one). This allowed us to make a box to box comparison where the
contribution to the power consumption for the empty chassis is similar and the
volume occupied in a rack is exactly the same. In particular in our test setup
we had 5 Atom cartridges and 15 X-Gene1 ones. We measured the throughput of
ParFullCMS by running it on an increasing number of cartridges, while at the
same time measuring the power consumption of the whole box. In figure
~\ref{fig:box2box} you have the results for our measurements in form of blue
dots and red crosses. Since the two boxes are filled with a different number of
cartridges, we cannot compare the two dataset directly, as the power cost due to
idle cartridges is different. What we did instead is to used the data we
acquired to evaluate the power consumption of a cartridge in running and idle
mode and then use those values to project the results for fully populated boxes.
The results are extremely unsatisfactory for the X-Gene1 based m400 cartridge
which seems to be extremely underperforming in terms of power consumption. We
attribute this to actual maturity issues of the cartridge itself, which
suspiciously has the same power consumption of the development board, rather
than actual power consumption of the CPU. This well illustrates the fact that
while promising, a lot needs to be done in term of production readiness for
ARMv8 based solution.

\section{Conclusions}
We continued our investigation effort in evaluating alternative platforms for
HEP workloads. In particular we have extended our analysis to include more
benchmarks and novel platforms. While not final, our conclusion shows that
depending on the various operating constrains in terms of power usage and
performance, either Atom based systems or latest generation Intel Xeon system
provide the best power efficiency levels. While APM X-Gene platform is still
relevant, it's clear that in order to compete with Intel, ARM based solutions
need not only to develop a performing CPU and match Intel pace in evolving their
products, but they need to address the maturity issues of their ecosystems, both
in terms of software and of auxiliary electronics. From the mere performance
point of view, both the POWER8 and even more so Xeon Phi results show how
difficult it is to take advantage of extremely parallel architecture without
specifically design software around them.

\section*{Acknowledgements}
This work was partially supported by the National Science Foundation, under
Cooperative Agreement PHY-1120138, and by the U.S. Department of Energy. We
would like to express our gratitude to APM for providing hardware and effort
benchmarking Geant4 ParFullCMS, to Intel for providing and managing some of the
Intel Xeon used, and to TechLab at CERN for providing and managing other Intel
Xeons used, the POWER8 system, the HP Moonshot systems and Intel Xeon Phi server
for the benchmarks.

\section*{References}
\bibliographystyle{unsrt}
\bibliography{chep2015-armv8-atom}

\end{document}